\documentclass[%
 sor,
 jor,
 amsmath,amssymb,
reprint,%
]{revtex4-1}

\usepackage{graphicx}
\usepackage{dcolumn}
\usepackage{bm}
\usepackage{color}
\definecolor {myc} {rgb} {0,0,0}  

\begin{document}


\title[Brittle fracture of polymer transient networks]{Brittle fracture of polymer transient networks}

\author{S. Arora}

\affiliation{Laboratoire Charles Coulomb, UMR 5521, CNRS, Universit\'e de Montpellier, F-34095, Montpellier CEDEX 05, France
}%

\author{A. Shabbir, O. Hassager}
\affiliation{%
Department of Chemical and Biochemical Engineering, Technical University of Denmark, 2800 Kongens Lyngby, Denmark
}%

\author{C. Ligoure, L. Ramos}
 \email{laurence.ramos@umontpellier.fr}
 \affiliation{%
Laboratoire Charles Coulomb, UMR 5521, CNRS, Universit\'e de Montpellier, F-34095, Montpellier CEDEX 05, France
}%

\date{\today}

\begin{abstract}
We study the fracture of reversible double transient networks, constituted of water  suspensions of entangled surfactant wormlike micelles  reversibly linked by various amounts of telechelic polymers. We provide a state diagram that delineates the regime of fracture without necking of the filament from the regime where no fracture or break-up has been observed. We show that filaments fracture when stretched at a rate larger than the inverse of the slowest relaxation time of the networks. We quantitatively demonstrate that dissipation processes are not relevant in our experimental conditions and that, depending on the density of nodes in the networks, fracture occurs in the linear {\color {myc} viscoelastic} regime or in a non-linear regime. In addition, analysis of the crack opening profiles indicates deviations from a parabolic shape close to the crack tip for weakly connected networks. We demonstrate
 a direct correlation between the amplitude of the deviation from the parabolic shape and the amount of non linear {\color {myc} viscoelasticity}.

%
\end{abstract}

\maketitle

\section{\label{sec:intro}Introduction}

How a material fails plays an important role in many fields ranging from material science to applied physics and geology. Capillarity is the relevant parameter to account for the break up of a liquid, as exemplified in the Rayleigh-plateau instability that drives a liquid filament to destabilize into finite size droplets~\cite{Eggers:2008hq}. Elasticity is obviously crucial to understand the fracture of a solid~\cite{Buehler:2010dpa}, but also of viscoelastic fluids~\cite{Ligoure:2013bd}. The fracture of a solid can be classified as brittle when no dissipative processes take place, or as ductile when dissipation comes into play. The behavior of viscoelastic samples is more complex as one may expect viscous dissipation to be enhanced and eventually dominate, thereby causing the sample to flow instead of break. However, when submitted to deformation rates larger than the inverse of their slowest relaxation times, viscoelastic fluids can break as solids do, as evidenced in several works with associative polymer networks~\cite{ZHAO:1993ja,BERRET:2001be,Tripathi2006, Tabuteau2009,Skrzeszewska:2010cpa, Tabuteau2011, Foyart2013}, entangled wormlike micelles~\cite{Rothstein2003, Smolka2003, Handzy2004, Chen2004, Bhardwaj2007}, networks combining wormlike micelles and associative polymers~\cite{Foyart2016}, polymer melts above the glass transition~\cite{Ide1977,Malkin:1997hx,Wang:2007kh, Huang2016, Huang:2016aa,Shabbir2016} or concentrated colloidal suspensions~\cite{Smith2015}.

Several challenges have however to be faced when dealing with viscoelastic samples.
From a conceptual point of view, the extension of the theoretical models for the fracture of solids to viscoelastic liquids, and of the definition of brittleness and ductility of solids to viscoelastic liquids, are not trivial. Moreover, from an experimental point of view, because viscoelastic samples flow, standard tools to investigate the fracture of solids cannot generally be used. Novel geometries have therefore been considered to investigate fracture processes in viscoelastic fluids, such as pendant drop experiments~\cite{Smolka2003, Tabuteau2009, Tabuteau2011}, using a shear cell~\cite{BERRET:2001be,Tabuteau2009,Skrzeszewska:2010cpa}, a Hele-Shaw cell and forcing a liquid to invade rapidly a thin layer of viscoelastic fluid~\cite{ZHAO:1993ja,Lefranc2014, Foyart2013, Foyart2016} or more complex geometries~\cite{Smolka2003, Chen2004, Handzy2004, Chakrabarti2016}. Filament stretching rheometry~\cite{Mckinley2002} appears as an exquisite tool to investigate the fracture of viscoelastic fluids, since it allows a sample to be submitted to a prescribed constant extensional rate. Filament stretching rheometry was successfully used to measure the tensile stress of associative polymer networks and wormlike micelles before fracture~\cite{Rothstein2003, Bhardwaj2007, Tripathi2006}. A substantial improvement has been recently achieved by coupling filament stretching rheometry to a fast imaging of the filament, allowing one not only to visualize but also to quantify the crack nucleation and propagation~\cite{Huang:2016aa}.

In this paper, we use filament stretching rheometry coupled to fast imaging to investigate the fracture processes of self-assembled polymer networks of controlled and tunable structure and viscoelasticity. We show, for the first time to the best of our knowledge in transient networks, a departure of the crack opening profile from the parabolic shape expected from a linear elastic fracture mechanism. We experimentally evidence the crucial role of non-linearities in the fracture of transient networks and demonstrate, using networks with a tunable connectivity, a direct correlation between non-linear {\color {myc} viscoelasticity} and departure from a parabolic profile.

The paper is organized as follows. We first present the linear shear viscoelasticity and the extensional rheology of the polymer networks. The modes of deformation up to a maximum strain are then described. We focus on the cracks that occur when the samples is strained at a sufficiently high rate and characterize their opening profiles and propagation velocity. Finally, we discuss the correlation between the extensional rheology and the crack opening profiles, in light of the literature and of the specificity of our samples.

\section{\label{sec:MatMet}Materials and methods}

\subsection{\label{sec:materials}Materials}

We investigate self-assembled transient networks consisting of a semi-dilute solution of surfactant wormlike micelles eventually reversibly cross-linked by telechelic polymers {\color {myc} whose structure and shear rheology have been previously investigated by some of us}~\cite{Ramos2007, Nakaya-Yaegashi2008, Tixier2010, Ramos2011}. We use a mixture of cetylpyridinium chloride (CpCl) and sodium salicylate (NaSal) with a NaSal/CpCl molar ratio of $0.5$ dispersed in brine ($0.5$ M NaCl). This mixture is known to form long and flexible surfactant cylindrical micelles~\cite{Rehage1988, Berret1993}. The sample eventually comprises home-synthesized triblock telechelic polymers, which are made of a poly(ethylene oxide) (PEO) hydrophillic backbone (molecular weight $10,000$ $\rm{gmol^{-1}}$) with
hydrophobic aliphatic chains grafted at both ends, $C_nH_{2n+1}$, where $n$ = $23$. The samples are prepared by weight and are characterized by two parameters, the mass fraction of micelles $\varphi$ = $(m_{\rm{CpCl}}+m_{\rm{NaSal}})/m_{\rm{total}}$ and the amount of polymer $\beta$ = $m_{\rm{polymer}}/(m_{\rm{CpCl}} +m_{\rm{NaSal}})$. Here $m_{\rm{CpCl}}$, $m_{\rm{NaSal}}$, and $m_{\rm{polymer}}$ are respectively the mass of CpCl, NaSal and polymer, and $m_{\rm{total}}$ is the total mass of the sample. The amount of polymer $\beta$ is varied between $0$ and $55\%$ and $\varphi$ is fixed at $10\%$,  {\color {myc} corresponding to CpCl, respectively NaSal, molar concentrations of $230$ mM, respectively $165$ mM.}

\subsection{\label{sec:methods}Methods}

\subsubsection{\label{sec:shear}Shear rheology}

The sample linear viscoelasticity is investigated with standard rheometry, using  a stress-controlled rheometer (MCR 502 from Anton-Paar) equipped with a Couette geometry. Temperature is fixed at $T=25^\circ$C for all rheology measurements. The storage ($G'$) and loss ($G''$) moduli are measured for frequency in the range $(0.1-100)$ rad/s at a fixed and small amplitude strain ($\gamma=10\%$), ensuring that data are acquired in the linear regime.

{\color {myc} The viscoelasticity of comparable samples have been previously investigated and discussed in details~\cite{Nakaya-Yaegashi2008}. Here, the sample composition is varied to tune their viscoelasticity.} A simple wormlike micelles solution (without telechelic polymer, $\beta=0$) behaves as a pure Maxwell fluid and is characterized by a shear plateau modulus $G_0$ and a unique characteristic relaxation time $\tau$. $G_0$ is related to the mesh size of the network of wormlike micelles, and $\tau$ is the geometric mean of the characteristic time for breaking/recombination and the characteristic time for reptation~\cite{Cates1987}. These parameters are obtained by fitting the frequency dependence of the storage, $G'$ and loss modulus $G"$ with their theoretical expressions for a Maxwell fluid:

\begin{equation}
\label{eqn:G'}
G'(\omega) = \dfrac{G_0{(\omega\tau)}^2}{1+ {(\omega\tau)}^2}
\end{equation}

\begin{equation}
\label{eqn:G''}
G''(\omega) = \dfrac{G_0{(\omega\tau)}}{1+ {(\omega\tau)}^2}
\end{equation}

Samples that comprise telechelic polymers behave as two-mode Maxwell fluids, resulting from the coexistence of two coupled
networks~\cite{Nakaya-Yaegashi2008}, one related to the bridging of the micelles by the telechelic polymers (elastic plateau $G_{\rm{fast}}$, relaxation time $\tau_{\rm{fast}}$) and one related to the
micelle entanglement (elastic plateau $G_{\rm{slow}}$, relaxation time $\tau_{\rm{slow}}$ with $\tau_{\rm{slow}} > \tau_{\rm{fast}}$ ). For a two-mode Maxwell model, $G'(\omega)$ and $G''(\omega)$ read:

\begin{equation}
\label{eqn:2G'}
G'(\omega) = \dfrac{G_{\rm{slow}}{(\omega\tau_{\rm{slow}})}^2}{1+ {(\omega\tau_{\rm{slow}})}^2} + \dfrac{G_{\rm{fast}}{(\omega\tau_{\rm{fast}})}^2}{1+ {(\omega\tau_{\rm{fast}})}^2}
\end{equation}

\begin{equation}
\label{eqn:2G''}
G''(\omega) = \dfrac{G_{\rm{slow}}{(\omega\tau_{\rm{slow}})}}{1+ {(\omega\tau_{\rm{slow}})}^2} + \dfrac{G_{\rm{fast}}{(\omega\tau_{\rm{fast}})}}{1+ {(\omega\tau_{\rm{fast}})}^2}
\end{equation}

Here, the elastic plateau modulus is the sum of the elastic moduli of the fast and slow modes, $G_0 = G_{\rm{slow}} + G_{\rm{fast}}$.

\subsubsection{\label{sec:extensional}Extensional rheology coupled to imaging}

To measure the mechanical response of the samples under an extensional deformation flow field, a VADER 1000 (versatile accurate deformation extensional rheometer) from Rheo Filament ApS~\cite{Huang2016} is used. The sample is loaded between two vertically aligned cylindrical stainless steel plates of identical diameters ($6$ mm or $9$ mm). The top plate is moved upwards leading to a stretching of the filament. The Hencky strain $\varepsilon$ is defined as $\varepsilon=-2 \ln(D(t)/D_0)$, where $D_0$ is the initial diameter (typically $D_0 \approx 2.5$ mm  in our experiments) and $D(t)$ is the mid-filament diameter at time $t$ during the stretching, as measured using a laser micrometer. The motion of the top plate is governed by a control loop scheme, with an active feedback control, to ensure a constant Hencky strain rate, $\dot{\varepsilon}=\frac{\partial \varepsilon}{\partial t}$~\cite{Marin2013} and a uniaxial stretching at the mid-filament plane.
The extensional stress $\sigma$ is defined as the mean stress difference at the mid-plane filament and is related to the measured force $F$ as $\sigma= \frac{F}{(\pi/4)D^2}$. One defines the transient extensional viscosity, or tensile stress growth coefficient, as $\eta_E^+ = \frac{\sigma}{\dot{\varepsilon}}$.
For a one-mode Maxwell fluid, linear viscoelasticity predicts the transient extensional viscosity, or tensile stress growth coefficient, to read

\begin{equation}
\label{eqn:etaE1mode}
{\eta_E}^+ = 3 G_0 \tau [1-\exp(-t/\tau)]
\end{equation}

For a two-mode Maxwell fluid~\cite{barnes1989},

\begin{equation}
\label{eqn:etaE2mode}
{\eta_E}^+ = 3 \bigl \{G_{\rm{slow}} \tau_{\rm{slow}} [1-\exp(-t/\tau_{\rm{slow}})]+ G_{\rm{fast}} \tau_{\rm{fast}} [1-\exp(-t/\tau_{\rm{fast}})] \bigr \}
\end{equation}

In our experiments, different strain rates, in the range $(0.03-2)$ $\rm{s^{-1}}$, are applied to measure the response of the viscoelastic materials in the viscous and elastic regimes. The maximum Hencky strain reported is $\varepsilon_{\rm{max}}\approx 3$. The main reason for this limitation is noise in the force signal for values less than about $10^{-3}$N (corresponding to $0.1$ g). All experiments are performed at $T \approx 25^\circ$C.

The imaging of a filament during its stretching is performed using a high speed camera (Photron Mini UX100) coupled with the VADER 1000 rheometer \cite{Huang2016b}. Because the active feedback control cannot be used concomitantly with the imaging, the kinematic trajectory of the top plate is defined by feedforward control parameters~\cite{Marin2013}. These parameters are obtained from experiments (without imaging) performed on VADER in a control loop scheme with an active feedback control for a given strain rate. Time series of filament stretching is recorded at the operating setting of $5000$ frames/s with a resolution of $(1280 \times 1000)$ $\rm{pixel^2}$.

\subsubsection{\label{sec:imageanalysis}Image analysis}

Images are analyzed with Matlab. To quantify the opening profile of cracks, we have developed a code that picks the coordinates marking the outline of the crack. The code is based on intensity cut-off value, which allows one to pick intensity values on the periphery of the crack. This way, one can compute the corresponding position coordinates and the vertex of the crack tip.

\section{\label{sec:LinearRheo}Linear viscoelasticity}

\begin{figure}
\includegraphics[width=1\columnwidth]{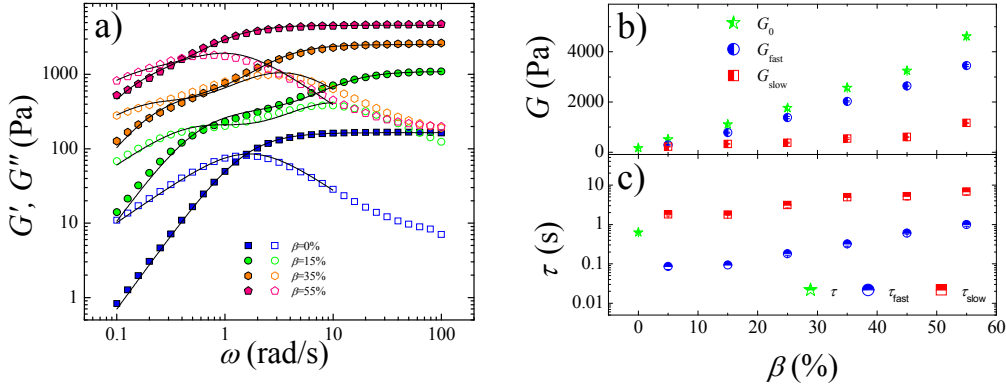}
\caption{\label{fig:fig1} Linear shear rheology for samples with various polymer concentrations, $\beta$, as indicated in the legend. (a) Evolution of the strorage and loss moduli with frequency. The symbols are the experimental data points and the continuous lines are the fit with one-mode (for $\beta=0$) and two-mode (for $\beta\neq0$) Maxwell fluid models. (b,c) Fit parameters, elastic moduli (b) and characteristic relaxation times (c), as a function of $\beta$.}
\end{figure}

Figure~\ref{fig:fig1}a shows the frequency dependence of $G'$ and $G''$ together with the Maxwell fits (one-mode Maxwell fluid for pure wormlike micelles, $\beta=0$, and two-mode Maxwell fluid otherwise), which account very well for the experimental data. Figure~\ref{fig:fig1}b summarizes the evolution with the amount of telechelic polymer, $\beta$, of the rheological characteristics of the samples as extracted from the fits. We measure that the shear plateau modulus, $G_0$, increases monotonically with the amount of telechelic polymer, $\beta$, from $170$ Pa in the absence of polymer to $4600$ Pa for $\beta=55$ \%. Similarly, we measure that both relaxation times continuously increase with $\beta$. The characteristic relaxation time $\tau_{\rm{slow}}$ of the micelles network increases from $0.6$ s without polymer up to $7$ s, for $\beta=55$ \%. The faster relaxation time, related to the network of telechelic polymers, is systematically about one order of magnitude smaller than the slower one. Note that, although the slow mode is related to the network of wormlike micelles, its rheological characteristics ($G_{\rm{slow}}$, $\tau_{\rm{slow}}$) are measured to continuously increase with the amount of telechelic polymer, demonstrating a coupling between the two networks.

\begin{figure}
\includegraphics[width=1\columnwidth]{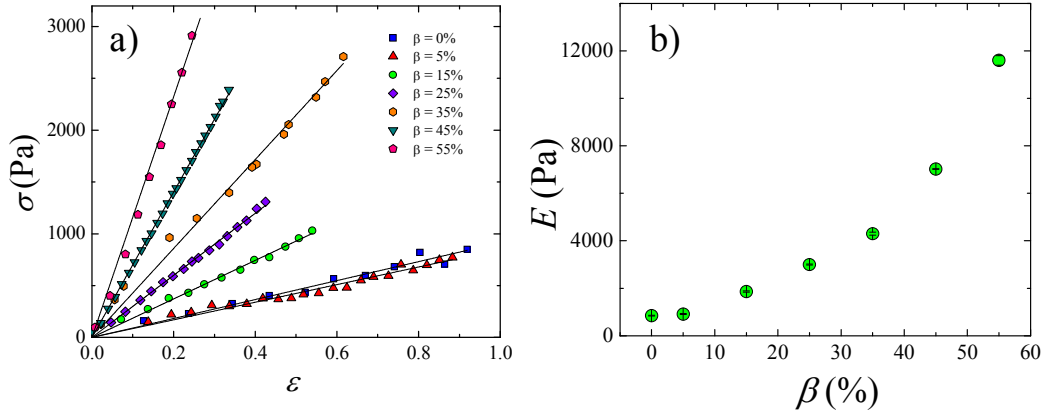}
\caption{\label{fig:fig2}  (a) Extensional stress as a function of the Hencky strain, $\varepsilon$, in the linear regime (small $\varepsilon$), for samples with various polymer concentrations, $\beta$, as indicated in the legend. The symbols are experimental data points and the lines are linear fits whose proportionality constant is the elastic modulus, $E$. Data have been acquired in the elastic regime (with $Wi$ in the range (1-4) depending on the samples). (b) $E$, as a function of $\beta$. }
\end{figure}

On the other hand, the Young modulus can be measured using extensional rheology. This requires data to be acquired at a sufficiently large extensional rate $\dot{\varepsilon}$ as compared to the slowest relaxation time so that viscous dissipation is not relevant. Figure~\ref{fig:fig2}a shows the growth of the measured stress $\sigma$ as a function of the strain $\varepsilon=\dot{\varepsilon} t$, with $\dot{\varepsilon}$ the imposed extension rate and $t$  the time elapsed since the sample is strained. In the limit of small deformation, $\sigma$ is measured to be proportional to $\varepsilon$, as expected. {\color {myc} Indeed, in the short time regime, i. e. for $t\ll\tau_{\rm{slow}},\tau_{\rm{fast}}$, Eq.~\ref{eqn:etaE2mode}
 reduces to $\eta_E^+ = \frac{\sigma}{\dot{\varepsilon}}=3 (G_{\rm{slow}}+G_{\rm{fast}})$, hence $\sigma=E\varepsilon$ where the proportionality constant is the Young modulus $E=3 (G_{\rm{slow}}+_{\rm{fast}})=3 G_0$ as expected  for an isotropic incompressible material. By fitting the experimental data at small deformation, the Young modulus is measured. We find that $E$ increases from $850$ to $11600$ Pa as $\beta$ increases (fig.~\ref{fig:fig2}b), and is comparable to $3$ times the shear modulus measured independently (fig.~\ref{fig:fig1}b).}

\section{\label{sec:Ext}Modes of rupture of the filaments}

\subsection{\label{sec:scheme}State diagram}

\begin{figure}
\includegraphics[width=1\columnwidth]{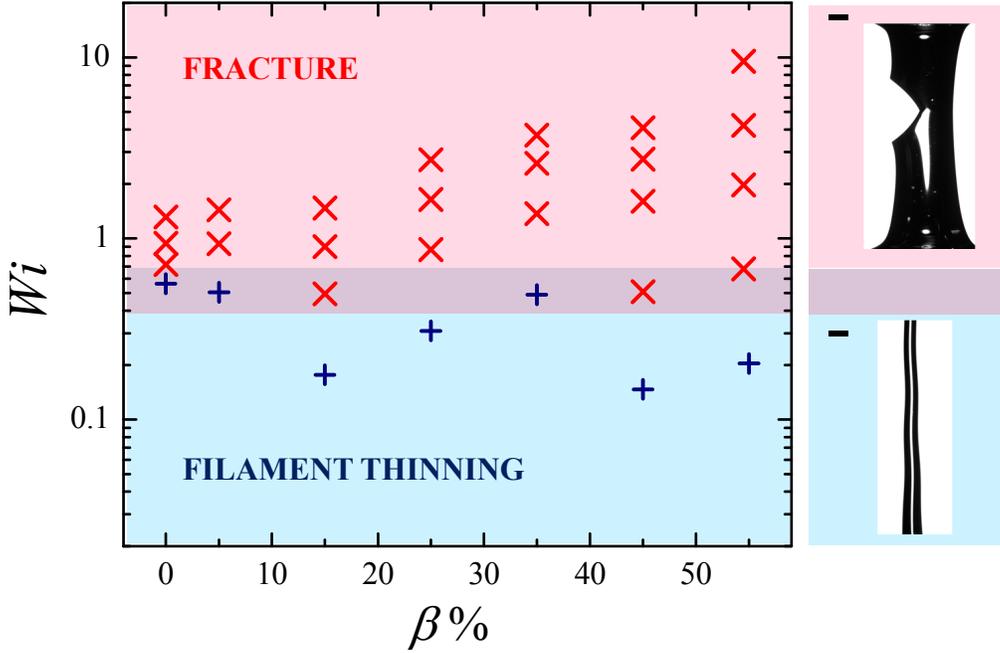}
\caption{\label{fig:fig3} Schematic state diagram, Weissenberg number, $Wi$, as a function of the amount of telechelic polymers in the sample. The symbols correspond to the experimental configurations investigated. The two regimes, continuous thinning at low $Wi$, and fracturing without necking at high $Wi$, are shown. Representative images of the filament corresponding to the two regimes are displayed.  For the top image (crack) crack, $\beta=25$ \% and $\dot{\varepsilon}=1 \, \rm{s}^{-1}$, and for the bottom image (capillary thinning), $\beta=0$ and $\dot{\varepsilon}=1 \, \rm{s}^{-1}$. Scale bars: $1$ mm.}
\end{figure}

We perform extensional rheology measurements using the feedback loop to ensure constant extensional rates, $\dot{\varepsilon}$, for samples with various copolymer contents, $\beta$. Because samples differ by their characteristic relaxation times (as shown in fig.~\ref{fig:fig1}c), the relevant quantity is not $\dot{\varepsilon}$ but the Weissenberg number, defined as $Wi=\dot{\epsilon}\tau$ for one-mode Maxwell fluid and as $Wi=\dot{\varepsilon}\tau_{\rm{slow}}$, for two-mode Maxwell fluids. In our experiments, $Wi$ is varied over about two orders of magnitude (from $0.1$ to $9.5$).
Visualization of the filament during its extension indicates two distinct types of material behavior, either a continuous liquid-like thinning up to the maximum Hencky strain $\varepsilon_{\rm{max}}\approx 3$ or by solid-like fracturing without necking~\cite{Tripathi2006, Tabuteau2009} at a Hencky strain below $\varepsilon_{\rm{max}}$. {\color {myc} Note that elastocapillary  break-up~\cite{Eggers1997, Anna2001} is not observed for Hencky strains less than 3. Clearly the thinning cannot continue to arbitrarily large Hencky strains, since capillary forces must ultimately dominate as the radius tends to zero. However the resulting capillary up is not the object of the present study. For all experimental conditions therefore, the behavior of the filament under a constant elongation rate is categorized into one of these two classes, liquid-like (thinning) or solid-like (fracture).} All data are reported in a schematic state diagram (fig.~\ref{fig:fig3}) where $Wi$ is plotted as a function of $\beta$. We observe that the filament thins continuously at low $Wi$ and cracks at a finite thickness (without necking) at higher $Wi$. The transition from continuous flow to filament rupture is similar to previous experimental observations on viscoelastic solutions of wormlike micelle~\cite{Bhardwaj2007} and associating polymer~\cite{Tripathi2006}. Remarkably, for all samples investigated here, the transition from thinning to fracturing occurs for a comparable critical Weissenberg number $Wi_c \approx (0.5-0.6)$. This experimental result follows remarkably well the theoretical Weissenberg criterion~\cite{Malkin:1997hx} for the rupture of polymeric liquids in extension with a constant strain rate that predicts $Wi_c=0.5$ for an elastic (upper convected Maxwell) liquid.

\subsection{\label{sec:ExtRheo}Non-linear extensional rheology}

\begin{figure}
\includegraphics[width=1\columnwidth]{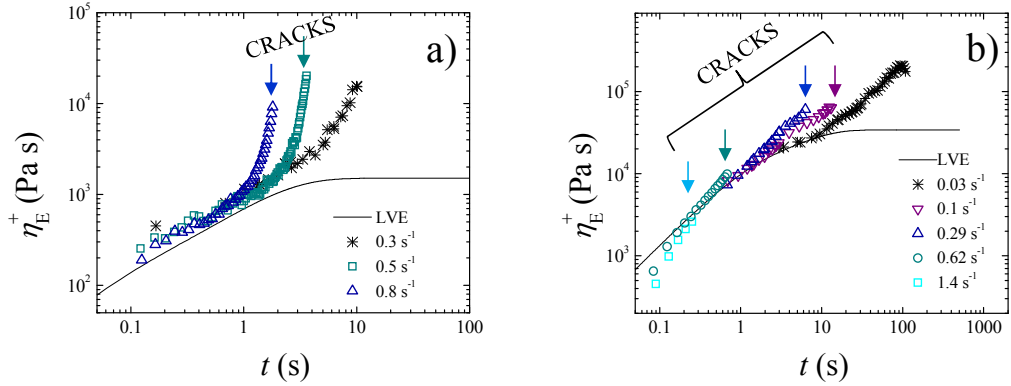}
\caption{\label{fig:fig4} Extensional stress growth coefficient, for two samples with $\beta=5$ \% (a) and $\beta=55$ \% (b), and for various extensional rates, as indicated in the legends. The symbols are experimental data points and the continuous lines are the theoretical expectations for linear viscoelasticity. Black crosses correspond to filament thinning and colored symbols correspond to cracks. The arrows indicate the time when fracture occurs.}
\end{figure}

To assess the importance of non-linearity, we compare the time dependence of the transient extensional viscosity, or tensile stress growth coefficient, $\eta_E^+$, to the linear viscoelasticity expectations. As an illustration, data acquired at different extensional rates for two samples differing by their amount of copolymer, $\beta=5$ \% (fig.~\ref{fig:fig4}a) and $\beta=55$ \% (fig.~\ref{fig:fig4}b), are displayed together with the linear viscoelastic expectations computed thanks to the linear viscoelastic parameters determined using a shear rheometer. For a two-mode Maxwell fluid, linear viscoelasticity predicts that $\eta_E^+$ continuously increases with time until reaching a plateau for time larger than the inverse of the slowest relaxation time, $\tau_{\rm{slow}}$ (Eq.~\ref{eqn:etaE2mode}). We measure that, at early times, the extensional rheology data follow the expected linear viscoelastic behavior. The nice quantitative agreement ensures the reliability of the two sets of measurements. In addition, we measure that crack (colored symbols) occurs at a time $t_c$ that decreases as the imposed extensional rate $\dot{\varepsilon}$ increases, such that the cumulated strain experienced by the sample, $\dot{\varepsilon} \, t_c$, is roughly constant (of the order of $1$). The marked difference between the two samples lies in the fact that for one sample the fracture occurs nearly in the linear regime ($\beta=55$ \%, fig.~\ref{fig:fig4}b), whereas the other sample exhibits a significant departure from the linear behavior before fracturing ($\beta=5$ \%, fig.~\ref{fig:fig4}a).

\begin{figure}
\includegraphics[width=1\columnwidth]{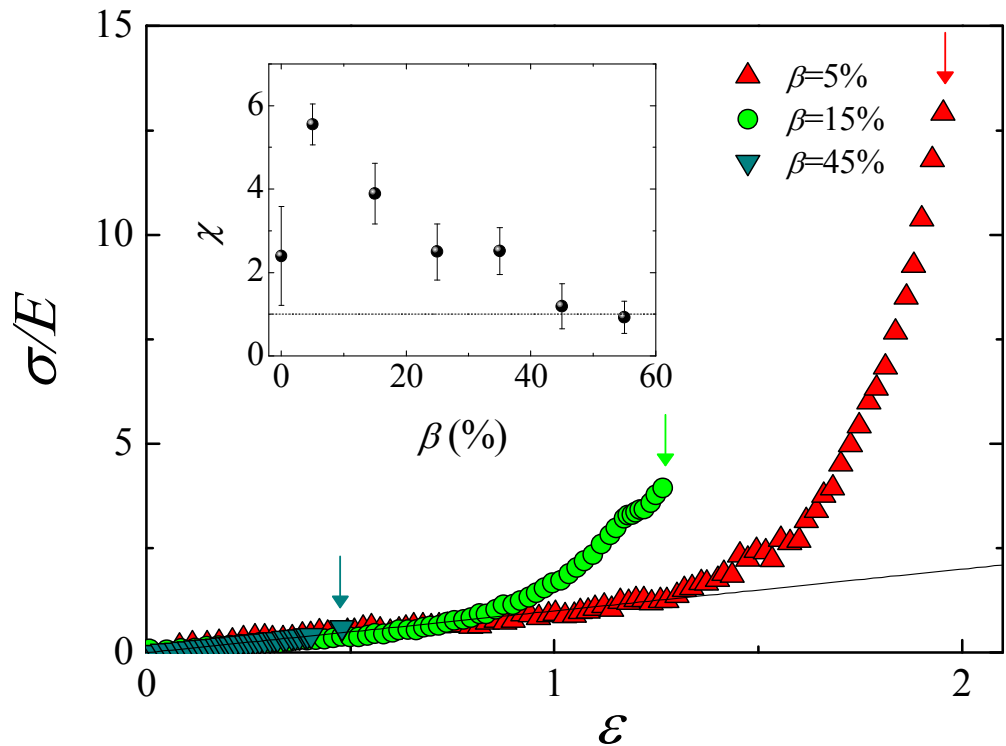}
\caption{\label{fig:fig5} Extensional stress normalized by the Young modulus measured in the linear regime at low strain (fig.~\ref{fig:fig2}a) as a function of the strain, for samples with different amounts of telechelic polymers, $\beta$, as indicated in the legend. For all samples, the strain rates have been adjusted such that $Wi\sim1$. The symbols are the experimental data points and the continuous line is the linear regime expectation $\sigma / E = \varepsilon$.  Fracture occurs at $\sigma_c$  and $\varepsilon_c$. Inset: Variation of $\chi=\frac{\sigma_c}{E \varepsilon_c}$ with $\beta$.  The dotted line indicates the linear {\color {myc} viscoelastic} regime.}
\end{figure}

The crucial role of the sample structure on the non linear viscoelastic behavior is also clearly seen in fig.~\ref{fig:fig5}, where data for samples with different amounts of telechelic polymers are plotted together. To account for the varying elasticities, the measured stress, $\sigma$, is normalized by the modulus, $E$, as measured in the {\color {myc} short time} regime, at small strain $\varepsilon$. By definition, data collapse at small strains. Although all samples crack at a similar strain of the order of $1$, they do so in very different manner: the sample with a large amount of polymer seems to fracture more or less in the linear {\color {myc} viscoelastic} regime whereas the sample comprising a small quantity of polymer exhibits large deviation from the linear regime before fracturing. More quantitatively, we measure the stress, $\sigma_c$,  and the strain, $\varepsilon_c$, at which the sample fractures. For a given sample, we find that, in the range of extensional rates $\dot{\varepsilon}$ investigated, $\varepsilon_c$ and $\sigma_c$ only weakly depend on $\dot{\varepsilon}$, as previously observed for pure wormlike micelles solutions~\cite{Rothstein2003, Bhardwaj2007}.  We define $\chi=\frac{\sigma_c}{E \varepsilon_c}$, which quantifies non linearity. For a sample that breaks in the linear {\color {myc} viscoelastic} regime, one expects $\sigma_c=E \varepsilon_c$, hence $\chi=1$. For a sample that strain-hardens, $\chi>1$. We report in the inset of fig.~\ref{fig:fig5} the evolution of $\chi$ with the amount of telechelic polymer, $\beta$, where data acquired at different $\dot{\varepsilon}$ are averaged. We find that $\chi$ decreases as $\beta$ increases, from values larger than $5$ for sample with a low amount of polymer down to values of the order of $1$ for $\beta \geq 45$ \%.
Thus, weakly connected samples (i.e. sample containing low amount of telechelic polymers) exhibit significant strain-hardening before fracture, whereas more connected samples break in the linear {\color {myc} viscoelastic} regime. Note that the numerical value found here for a sample without telechelic polymer is in agreement with the one measured in~\cite{Bhardwaj2007} for a comparable system.

Hence, extensional rheology demonstrates that, although all samples displays qualitatively similar linear viscoelastic behavior (which can be well accounted by a Maxwell model), they present very different non-linear viscoelastic behaviors and fracture processes can occur in the linear regime  ($\chi\approx1$) or after a significant strain-hardening ($\chi>1$) depending on the sample connectivity.

\subsection{\label{sec:imaging}Crack imaging}

\begin{figure}
\includegraphics[width=1\columnwidth]{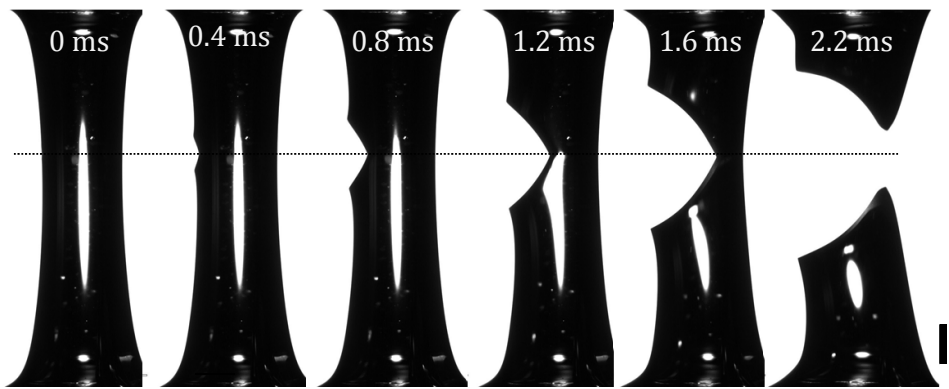}
\caption{\label{fig:fig6} Time series of a crack propagating in a sample with $\beta=25$ \% and strain $\dot{\varepsilon}=1\, \rm{s}^{-1}$. Scale bar: $1$ mm.}
\end{figure}

To better understand the links between the sample structure, the non-linear extensional rheology, and the fracture process, we use a fast camera to image the cracks during the extension of a filament at a prescribed rate.  Figure~\ref{fig:fig6} displays a representative time series during the crack propagation of a sample with $\beta=25$ \%. Cracks are imaged for various  samples with $\beta$ ranging from $15$ to $55$ \% . In all cases, a single crack forms that systematically propagates straight, perpendicularly to the extension direction.

\begin{figure}
\includegraphics[width=1\columnwidth]{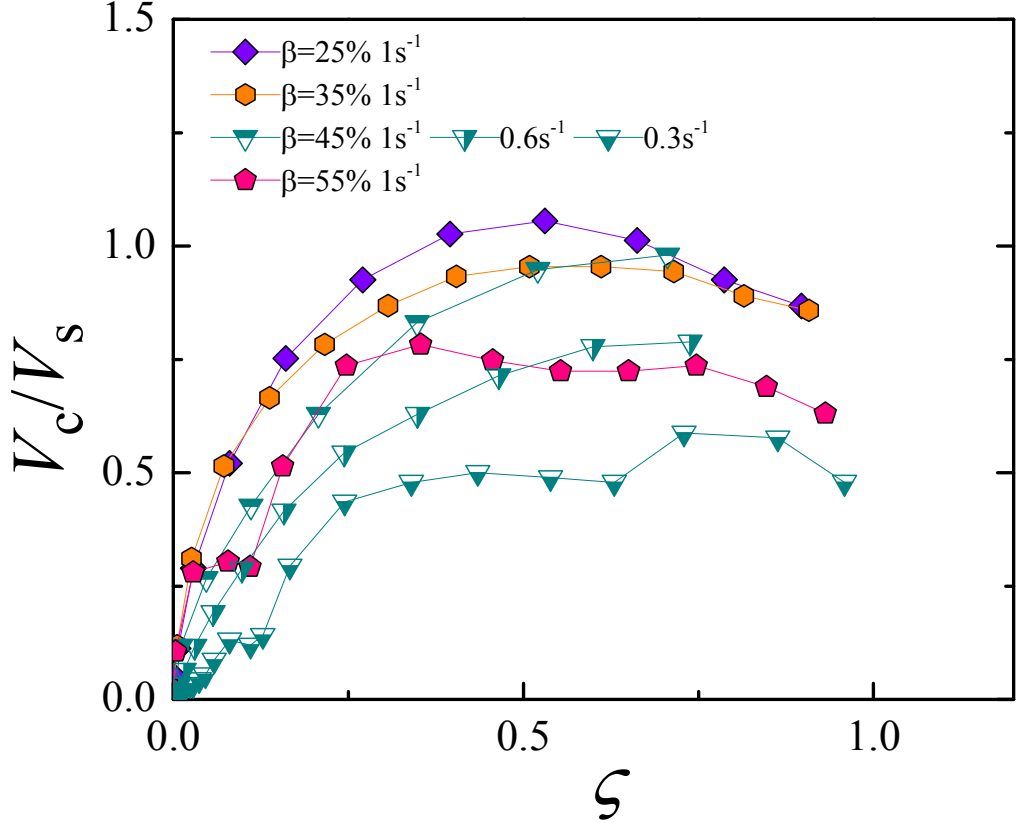}
\caption{\label{fig:fig7} Velocity of the crack tip as a function of the run distance, for samples with different amounts of telechelic polymers and different strain rates, as indicated in the legend. The velocity is normalized by the shear sound velocity and the distance by the total thickness of the filament, such that $\varsigma=0.5$ correspond to the mid-filament.}
\end{figure}

By tracking the crack tip, the instantaneous velocity of the crack, $V_c$, can be measured. To account for the various elastic moduli of the samples considered here, data have to be compared with the shear wave sound velocity, $V_s$, of each sample. For a solid of shear modulus $G_0$, $V_s=\sqrt{{G_0}/{\rho}}$, with $\rho$ the sample density. For the samples investigated here, $V_s$ varies between $1$ and $2$ m/s. Figure~\ref{fig:fig7} displays for all samples the crack velocity normalized by the sound velocity, $V_c/V_s$. $V_c/V_s$ increases steadily and reaches a plateau value more or less when the crack length become comparable to half the filament diameter ($\varsigma=0.5$). The cracks propagate fast as the steady state value is comparable to the shear wave velocity ($V_c/V_s$ ranges between $0.5$ and $1$). Finally, we mention that, although cracks propagate fast, they do not oscillate (as opposed to the finding of~\cite{Deegan2001, Livne2007}) probably because in our experiments cracks travel over very short distances.

\subsection{\label{sec:shape}Crack opening profiles}

\begin{figure}
\includegraphics[width=1\columnwidth]{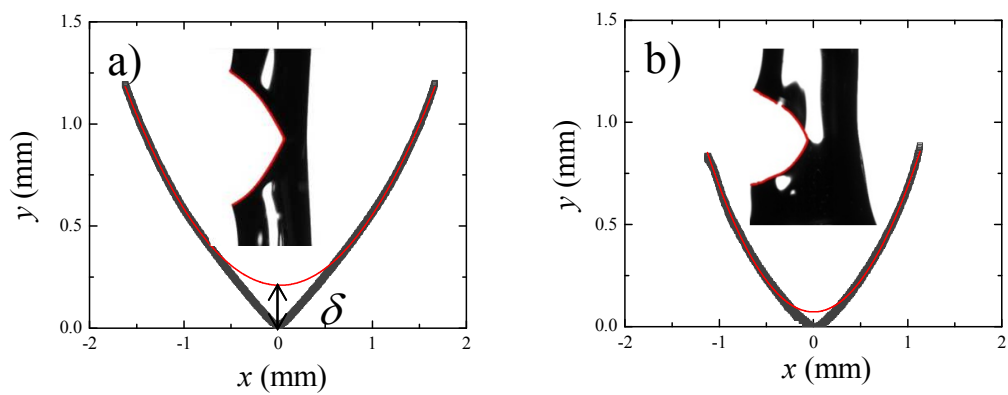}
\caption{\label{fig:fig8} Crack opening profiles measured when the cracks has propagated over a distance equal to half the filament thickness, for a sample with $\beta=15$ \% (a) and $55$ \% (b). The grey symbols are the experimental data points and the thin red lines correspond to the fit of the profile with a parabola, allowing the determination of the distance $\delta$.}
\end{figure}

An interesting feature concerns the crack opening profiles. As shown in fig.~\ref{fig:fig8}, the crack shape might depart significantly from the parabolic shape theoretically expected with the use of finite elasticity theory required for such soft materials \cite{Tabuteau2011}, and displays close to its tip a wedge profile.  The deviation from the parabolic shape is quantified by the length $\delta$ extracted from a fit of the profile of a crack (whose tip is positioned at $x=0$, $y=0$) with the functional form $y=\delta + ax^2$. The fit shown as red lines in fig.~\ref{fig:fig8} account very well for the experimental data and yield numerical values for $\delta$ between $0.017$ and $0.2$ mm, and for $a$ between $0.42$ and $0.76$ $\rm{mm^{-1}}$.

\begin{figure}
\includegraphics[width=1\columnwidth]{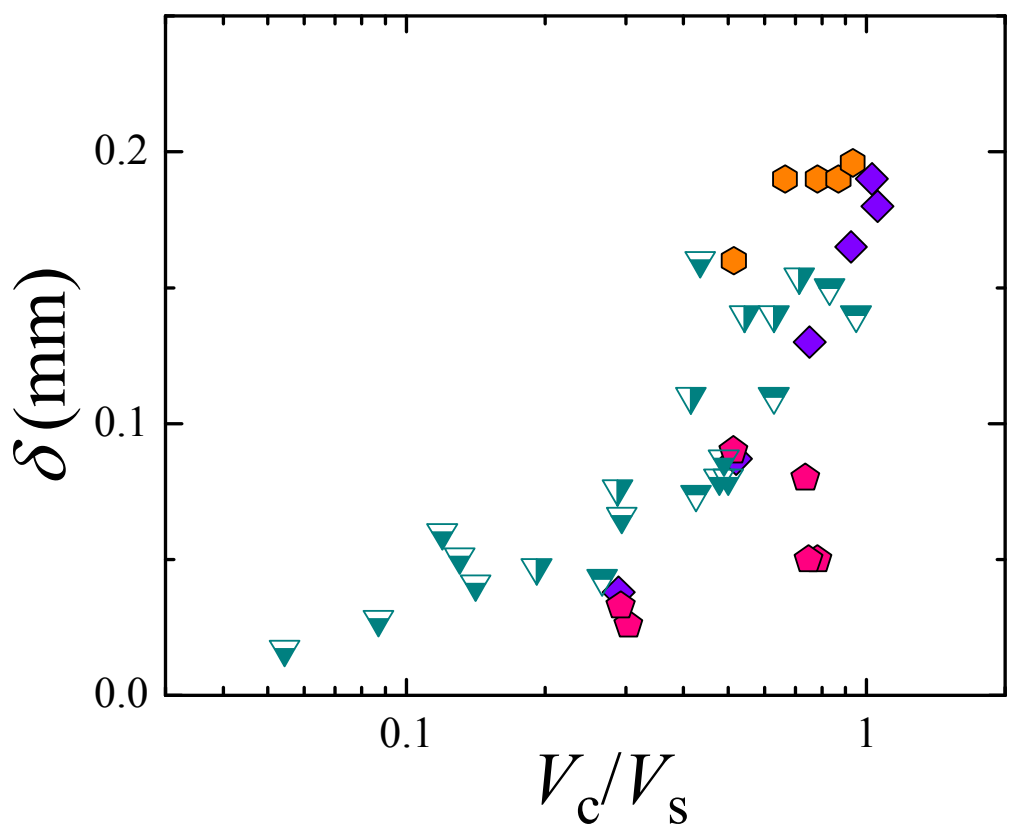}
\caption{\label{fig:fig9} Length $\delta$ as a function of the velocity of the crack tip normalized with the speed of the shear wave, for samples with different amounts of telechelic polymers and different strain rate. The symbols are the same as in fig.~\ref{fig:fig7}.}
\end{figure}

This type of profile has already been observed experimentally in elastomers and permanent gels~\cite{Deegan2001, Livne2008, Goldman2012, Morishita2016}. Similarly to previous works, we find that overall $\delta$ increases as the propagation velocity of the crack increases (fig.~\ref{fig:fig9}).

\section{\label{sec:Discussion}Discussion}

The samples investigated here are viscoelastic fluids, characterized by one (or two) relaxation times. We observe (fig.~\ref{fig:fig3}) that they fracture only if they are submitted to an extensional strain rate larger than the inverse of their longest relaxation time ($Wi \geq 1$). Hence we expect that for the large strain rates viscous dissipations are negligible. This is in accordance to the fact that the cracks propagate at very large speed, comparable to the shear wave velocity for a solid. More quantitatively, to check whether viscoelastic effects do play a role or not in the crack propagation, one has to compare the length over which the crack propagates $L$ and the length $\ell = V_c \tau_{\rm{slow}}$ over which viscoelastic effects are relevant~\cite{DeGennes1988, DeGennes1996}. In our experiments, the crack velocity in the steady state (at mid-filament) $V_c$ is typically of the order of $1$ m/s and the slow relaxation time of the network $\tau_{\rm{slow}}$ is in the range $(2-7)$ s. These values yield a length $\ell = V_c \tau_{\rm{slow}}$ that ranges between $2.5$ and $11$ m, while the radius of the filament, which dictates the length $L$ over which the crack propagates, is typically of $1$ mm. Hence in all cases, $\ell$ is several orders of magnitude larger than $L$, ensuring that viscous effects are not relevant. This is consistent with the fact that at large distance the crack is parabolic and does not exhibit the $x^{3/2}$ scaling predicted by  the viscoelastic trumpet model~\cite{DeGennes1996} and experimentally measured in the adhesive fracture of a polymer melt~\cite{Saulnier2004} and in polymer liquid under tension~\cite{Huang2016b}. In addition, because  the samples investigated here are very soft and strained elastically over very large deformation, they can be considered as hyperelastic. The importance of hyperelasticity in the vicinity of the crack tip may play an important role in the dynamics of fracture~\cite{Buehler2003, Buehler2006}. The characteristic size over which very large strains are involved close to the crack tip reads $R_{\rm{tip}}\approx \Gamma /G_0$ with $G_0$ the shear modulus and $\Gamma$ the fracture energy which can be approximated in the framework of finite elasticity as $\frac{\pi G_0}{4a}$~\cite{HUI:2003dv} (with $a$ the radius of curvature of the crack tip, extracted from the parabolic fit of the crack profile). Hence $R_{\rm{tip}}\approx 1/a$, and ranges in our experiments between $1.3$ and $2.4$ mm.  In addition to the criterium given above, the generalization of the visco-elastic trumpet model to hyperelastic  materials dictates  that $\frac{R_{\rm{tip}}}{L} \frac{L}{\ell}$ must be small~\cite{Tabuteau2011}  to ensure that viscous dissipations are negligible. Since in our experiment $R_{\rm{tip}}$ is comparable to $L$ and $L \ll \ell$, hence $\frac{R_{\rm{tip}}}{L} \frac{L}{\ell} \ll \ell$, and viscous relaxation are expected to be negligible. All these experimental facts show that filaments always fracture in the elastic limit and that dissipation is not relevant in the process. Hence, despite the samples are viscoelastic, the fractures are brittle in the experimental conditions considered here.

\begin{figure}
\includegraphics[width=1\columnwidth]{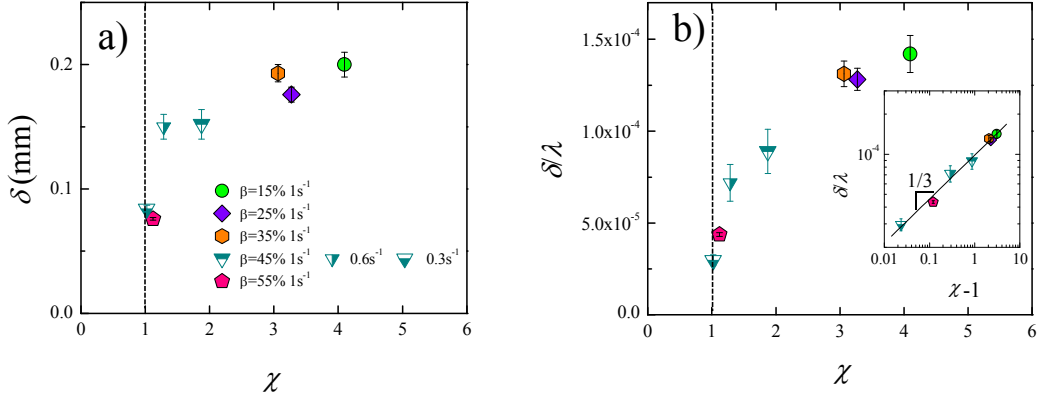}
\caption{\label{fig:fig10} Correlation between the non-linear elasticity quantified by
 $\chi=\frac{\sigma_c}{E \varepsilon_c}$  and the departure from the parabolic shape quantified by (a) the length $\delta$,  (b) the normalized length $\frac{\delta}{\lambda}$ (see text for a definition of $\lambda$). The dotted lines indicate linear {\color {myc} viscoelasticity}. Inset: normalized length as a function of $\chi-1$ in a log-log plot. The best fit yields a powerlaw with an exponent $1/3$.
}
\end{figure}

Our results can therefore be put in parallel with experiments on solid samples. The peculiar opening profile of the cracks that we measure has been previously observed, but only in elastomers~\cite{Deegan2001, Livne2008, Goldman2012,Morishita2016}. Livne et al. point out the importance of non linear elasticity close to the crack tip for the investigated soft incompressible elastomers (elastic moduli in the range $33-190$ kPa). They demonstrate that the length $\delta$ is not related to dissipative processes, i.e.  $\delta$ cannot be regarded as a characteristic length of a process zone, and argue indirectly that $\delta$ is related to finite elasticity. More recently, a correlation has been experimentally shown between $\delta$ and the hyperelasticity measured independently for elastomers filled with various amounts of carbon black~\cite{Morishita2016}.
Our data allows us to check for a direct correlation between the shape of the opening profile and the sample non linear {\color {myc} viscoelasticity}. As discussed above, non linear {\color {myc} viscoelasticity} can be quantified with $\chi=\frac{\sigma_c}{E \varepsilon_c}$, where  $\sigma_c$, resp. $\varepsilon_c$, is the stress, resp. the strain at which a crack nucleates and $E$ is the sample modulus. We show in fig.\ref{fig:fig10}a that $\delta$ is rather small (of the order of $0.08$ mm) when the samples cracks in the linear elastic{\color {myc} viscoelastic} regime ($\chi\approx1$) and continuously increases with  $\chi$, directly demonstrating the correlation between the amount of non linear elasticity and the departure from the parabolic shape of the crack profile. It is interesting to compare $\delta$ to a characteristic length of the fracture process. {\color {myc} Without crack, the radial speed for the contraction of the filament due to stretching is $V_{\rm{filament}}=R\dot{\varepsilon}$, with $R$ the radius of filament. One therefore proposes to define a characteristic length of the fracture process $\lambda$ as the radius of the filament times the ratio between the crack propagation speed and the natural speed of the filament in the absence of crack: $\lambda=R\frac{V_c}{V_{\rm{filament}}}=\frac{V_c}{\dot{\varepsilon}}$}. The plot of the non dimensional length $\frac{\delta}{\lambda}$ varies monotonically with $\chi$ (fig.\ref{fig:fig10}b). Intriguingly, we find that $\frac{\delta}{\lambda}$ varies as a powerlaw with $\chi -1$ with an exponent $1/3$, suggesting a critical phenomenon (inset fig.\ref{fig:fig10}b).

As a final remark, we wish to discuss our results in light of the sample structure. On the time and length scales considered here, the samples can be considered as blends of two coupled networks. One network is formed by the entangled wormlike micelles and the other one is formed by the telechelic polymers that link the wormlike micelles. We investigate a family of samples such that the network of wormlike micelles is kept constant and the density of network of telechelic polymers is varied, as $\beta$ changes. Our results show that double networks with a loose telechelic network strain hardens before fracturing. By contrast, when the telechelic polymer network is denser, the sample does not strain harden but fractures in the linear regime. Our findings suggest therefore that the capacity to strain harden is a specific feature of the wormlike micelles network, which may be impeded to do so due to the strong coupling with the telechelic network. This physical picture is consistent with experiments on various wormlike micelle systems. Indeed, for entangled wormlike micelles solutions, considerable strain hardening has been measured and modeled by the finitely extensible nonlinear elastic (FENE-PM) model which accounts for the finite extensibility of the wormlike micelles regarded as Gaussian chains~\cite{Walker1996, Rothstein2003}. Interestingly, Rothstein et al.~\cite{Rothstein2003, Bhardwaj2007} have also measured a decrease of the strain hardening when the concentration of wormlike increases, in agreement with our experimental observations, due to the increase of the density  of elastic nodes.

\section{\label{sec:Conclusion}Conclusion}

We have investigated the fracture processes in self-assembled double transient networks by combining fast imaging to a filament stretching rheometer. The networks break elastically without necking when deformed at a rate larger than the inverse of their lowest characteristic relaxation time. We have rationalized the non relevance of viscous dissipation effects  and from the analysis of the crack opening profile, we have evidenced for the first time for viscoelastic fluids and for the first time in  filament breaking experiments a departure at the  crack tip from the parabola expected from linear elasticity.  Thanks to the unique coupling in one single experiment between rheological measurements and crack opening profile characterization, we have provided a direct evidence of a correlation between the non-linear viscoelasticity and the shape of the crack profile close to the tip.  By varying the composition, we have tuned the sample viscoelasticity  and have found that samples with a dense networks of elastic nodes break in the linear regime whereas samples with a loose network of nodes exhibit significantly strain-hardening before fracturing.

\begin{acknowledgments}
This work was supported by the EU (Marie Sklodowska-Curie ITN Supolen, Grant No. 607937).
\end{acknowledgments}

%



\end{document}